\documentstyle[twoside,fleqn,epsf,espcrc2]{article}

\newcommand{\AmS}{{\protect\the\textfont2
  A\kern-.1667em\lower.5ex\hbox{M}\kern-.125emS}}

\title{Monopole clusters and critical dynamics in four-dimensional
U(1)\thanks{Work supported by the Deutsche Forschungsgemeinschaft under grant
No.~Schi 257/3-2 and Schi 257/1-4.}}
\author{A.~Bode, Th.~Lippert\thanks{Talk presented by Th.~Lippert.} and
K.~Schilling\\[8pt]
       {Department of Physics, University of Wuppertal,
        D-42097 Wuppertal, Germany}}%

\begin{document}

\begin{abstract}
We investigate monopoles in four-dimensional compact U(1) with Wilson
action.  We focus our attention on monopole clusters as they can be
identified unambiguously contrary to monopole loops.  We locate the
clusters and determine their properties near the U(1) phase
transition.  The Coulomb phase is characterized by several small
clusters, whereas in the confined phase the small clusters coalesce to
one large cluster filling up the whole system. We find that clusters
winding around the periodic lattice are absent within both phases
and during the transition.  However, within the confined phase, we
observe periodically closed monopole loops if cooling is applied.
\end{abstract}
\maketitle

\section{INTRODUCTION}
Ever since the work of DeGrand and Toussaint\cite{DEGRAND}, who gave a
prescription for constructing monopoles in compact QED, there is no
doubt that monopoles can yield a clear signal as to the location of
the U(1) phase transition in four dimensions and that one can
characterize the phase structure from the dynamical behavior of
monopoles. Their results have been confirmed by Barber who found a
strong correlation between the average plaquette action and the total
monopole length\cite{BARBER}.  Further evidence for the influence of
monopoles on the phase structure has been given in a recent work of
Bornyakov et al.~\cite{BORNYAKOV}. These authors suppressed
configurations containing monopoles, on the level of the partition
function.  This led to a deconfined phase throughout a
$\beta$-range from $0.1$ to $2$.

The latter result suggests that monopoles play an active role in
controlling the phase structure of U(1) rather than being just another
signal for the phase change. Such a scenario is supported by the
observation of very strong hysteresis effects\cite{GROESCH}, which
lead to supercritical slowing down (SCSD) for the U(1) updating.  As
to a possible mechanism how monopoles affect the phase structure and
handicap the efficiency of the updating, the authors of
Ref.~\cite{GROESCH} argued that periodically closed monopole current
loops, {\it i.~e.} loops closed due to the finiteness of the lattice
and the periodical boundary conditions imposed, are responsible for
long lived metastable states\footnote{Additionally, the authors traced
back small action gaps---deep in the weak coupling phase---to closed
Dirac-sheets.}.  It remains an open question, how in the infinite
volume limit---where periodically closed loops should be absent---SCSD
can be explained by such a mechanism. Or is SCSD just a finite-size
phenomenon in U(1)?

Our present approach is to proceed from previous {\it global} analyses
to an investigation of detailed {\it spatial} structures
 by locating closed monopole loops on the lattice and
determining their properties.

We have no well-founded
 criterion, however,  which would allow to identify single
closed loops unambiguously, in view of the observed proliferation
of  monopole line crossings. Therefore, we restrict our
considerations to {\it clusters} of loops, {\it i.~e.}~objects composed out of
connected monopole lines. We compute observables like the winding
number and the length of single clusters.
We will also try to uncover topological structures,
which might be hidden by  the local fluctuations.

\section{MONOPOLE LINES AND CLUSTERS}

\begin{figure}[htb]
\epsfxsize=75mm
\epsfbox{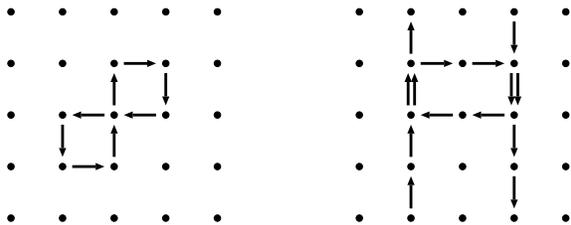}
\caption{Two examples of ambiguous loops.}
\label{AMBIGUOUS}
\end{figure}

We adopt the usual definition for monopole world lines in
four dimensions\cite{DEGRAND}:
\begin{equation}
m_{\mu}(x)=\frac{1}{2}\epsilon_{\mu\nu\rho\sigma}\left[
n_{\rho\sigma}(x+\nu)-n_{\rho\sigma}(x)\right],
\end{equation}
with
\begin{equation}
n_{\mu\nu}(x)=\frac{1}{2\pi}\left(
\overline{\Phi}_{\mu\nu}(x)-\Phi_{\mu\nu}(x)\right).
\end{equation}
The plaquette flux for the Wilson action varies between $-4\pi$ and
$4\pi$, the physical flux lies in the range between $-\pi$ and $\pi$,
and $n_{\mu\nu}$ can take on integer values between $-2$ and $2$.  The
divergence of the monopole links vanishes,
$\partial_{\mu}m_{\mu}(x)=0$, giving rise to closed loops of monopole
lines.  The monopole lines can intersect, and the identification of
single monopole loops is ambiguous as depicted in
Fig.~\ref{AMBIGUOUS}.
In the first example, either two plaquette-loops or one large loop
with one crossing can be identified.  The second example can either be
viewed as two periodically closed loops and one square loop or as one
large loop homotopic to a point.

We emphasize again
that the actual monopole clusters turn out to be utterly entangled.
Numerous loop crossings occur. Therefore, we have
concentrated our investigation on monopole clusters, {\em i.~e.}  objects
composed out of connected closed loops.

\section{RESULTS}
Our numerical work has been done on a connection machine CM-5.  We
designed data-parallel and message-passing algorithms in order to
locate the clusters\cite{LIPPERT1,LIPPERT2}.  We worked on lattices of
size $8^4$ and $16^4$.  For each size, we performed about 100000
sweeps using standard Metropolis updating and analyzed more than 10000
configurations.  In each case, we worked at the respective coexistence
point $\beta_c$\cite{LIPPERTLAT92}.

\subsection{Length distribution and loop crossings}

\begin{figure}[htb]
\epsfxsize=75mm
\vglue-3pt
\vglue-1.1cm
\epsfbox{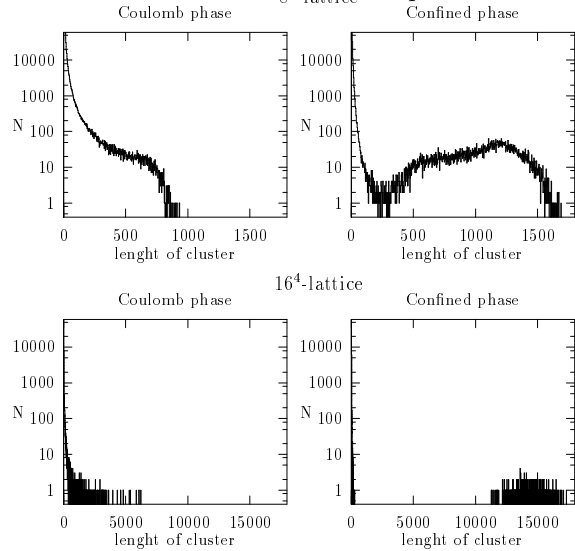}
\vglue 1.1cm
\vglue-3pt
\caption{Length distribution.}
\label{MONOPOLELENGTH}
\end{figure}

In Fig.~\ref{MONOPOLELENGTH}, we present the distribution of the
length of the clusters for both phases on an $8^4$- and a
$16^4$-lattice.
The confined phase is characterized by the occurrence of large
clusters. In particular, on the $16^4$-lattice the contribution of
smaller loops nearly vanishes and one cluster is filling up the whole
lattice.  In the Coulomb phase, the large cluster is split into
several smaller pieces. This goes along with a decreasing total length
of monopole lines. We interpret a tunnelling event from the Coulomb to
the confined phase as a condensation from a dilute monopole current
gas to a condensed cluster.  The simulation algorithm should be
capable to drive tunnelling, {\em i.~e.} to create and to destroy
large clusters.

\begin{figure}[htb]
\epsfxsize=75mm
\vglue-1.1cm
\vglue-6pt
\epsfbox{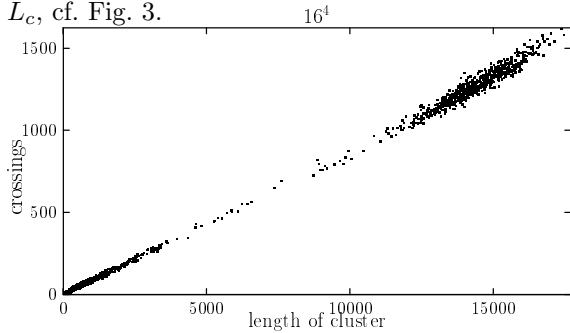}
\vglue-6pt
\vglue 1.1cm
\caption{Loop crossings on the $16^4$-lattice.}
\label{CROSSINGS}
\end{figure}

A four-dimensional visualization\cite{LIPPERT1} of the tunnelling
reveals an entangled jumble of monopole lines throughout the system.
In an attempt to look into those clusters, we analyzed the number of
crossings, $N_c$. We found nearly linear correlation between $N_c$ and
the length of the clusters, $L_c$, cf.~Fig.~\ref{CROSSINGS}.

This provides evidence
that small cluster condense into larger structures.

\subsection{Periodically closed clusters}

We define the {\em winding number} for the direction $\mu$ as
\begin{equation}
\label{winding}
W(\mu)=\frac{1}{L_{\mu}}\sum_{x\in \; \mbox{\tiny cluster}}m_{\mu}(x),
\end{equation}
where $L_{\mu}$ is the lattice extension in $\mu$-direction.  If this
number is non-zero the considered cluster is wrapping around the
lattice along the specified direction.

Among 10000 analyzed
configurations of  the $8^4$-lattice, we found only
8 events with  non-zero winding numbers.
These clusters came up within the flip region from the
confined to the Coulomb phase. The lifetimes of the
non-zero winding numbers ranged from one to two sweeps.
On the $16^4$-lattice, we did not encounter any periodically closed
cluster within the confined phase and in particular within the flip
region between the two phases.

These findings cast doubt on periodically closed monopole clusters being
responsible for long lived metastabilities at $\beta_c$.

\subsection{Cooling}

One might argue, however, that topologically important structures
within the monopole clusters might be hidden under
the local fluctuations.
In an attempt to remove the fluctuations and to reveal those
structures, we imposed two different types  of cooling on the
configurations, the first by minimizing the action and the second by
cooling adiabatically.

In the Coulomb phase, where the clusters are small, no non-zero winding
numbers appear through the cooling procedure.
Within the confined phase, however,
we observe that the cooling process gives rise
to the creation of such  winding
numbers.
The time history under cooling
depends strongly on the type of cooling,
generating winding numbers in different directions.
Furthermore, for a given  cooling procedure, the directions $\mu$,
Eq.~\ref{winding}, fluctuate
frequently from sweep to sweep.
This is another aspect of the fact that it is
impossible to identify single closed loops unambiguously.

Clearly, further investigations are needed
 in order to answer the question whether or not periodically closed
loops are an artifact of cooling.

\end{document}